\documentclass[12pt]{article}
\usepackage{amsfonts}
\usepackage{amssymb}
\usepackage{latexsym}
\usepackage{graphicx}
\usepackage[english]{babel}
\begin{document}
\input epsf

\renewcommand{\theequation}{\arabic{section}.\arabic{equation}}
\newcommand{\f}[2]{\frac{#1}{#2}}
\def\be{\begin{equation}}
\def\bea{\begin{eqnarray}}
\def\ee{\end{equation}}
\def\eea{\end{eqnarray}}
\def\c{\cosh\alpha}
\def\r{\rightarrow}
\def\pa{\partial}
\def\t{\tilde}
\def\n{\nonumber}
\def\nn{\nonumber\\ }
\def\g{\Gamma}
\def\h{{1\over 2}}
\def\ad{{\dot a}}
\def\b{\bigskip}
\def\m{\medskip}
\def\p{{\cal {P}}}
\def\q{{\cal {Q}}}
\def\qt{ {\tilde {\cal {Q}} }}
\def\s{{\cal {S}}}
\def\d{{d\over dt}}
\def\du{{d\over d\tau}}

\def\ph{\hat{\cal{P}}}
\def\qh{\hat{{\cal{Q}}}}
\def\qth{\hat{\tilde{\cal{Q}}}}
\def\sh{\hat{\cal{S}}}
\def\ih{\hat{I}}

\newcommand{\dt}[2]{\frac{d #1}{d #2}}

\begin{flushright}
\end{flushright}
\vspace{20mm}
\begin{center}
{\LARGE  Black hole size and phase space volumes}\\
\vspace{18mm}
{\bf  Samir D. Mathur\footnote{mathur@mps.ohio-state.edu}}\\
\vspace{8mm}
Department of Physics,\\ The Ohio State University,\\ Columbus,
OH 43210, USA\\
\vspace{4mm}
\end{center}
\vspace{10mm}
\thispagestyle{empty}
\begin{abstract}

For extremal black holes the fuzzball conjecture says that the  throat of the geometry ends in a quantum `fuzz', instead of being infinite in length with a horizon at the end. For the D1-D5 system we consider a family of  sub-ensembles of states, and find that  in each case the boundary area of the fuzzball satisfies a Bekenstein type relation with the entropy enclosed.  We  suggest a relation between the `capped throat' structure  of microstate geometries and the fact that the extremal hole was found to have  zero entropy in some gravity computations. We examine quantum corrections including string 1-loop effects  and check that they do not affect our leading order computations.

\end{abstract}
\newpage
\setcounter{page}{1}
\renewcommand{\theequation}{\arabic{section}.\arabic{equation}}
\section{Introduction}\label{introduction}
\setcounter{equation}{0}

It is a remarkable fact that the entropy of a black hole can be understood in terms of the surface area of its horizon \cite{bek}. Why is there such a connection between the size of the hole and its entropy? In this paper we make a small observation that might be relevant to this connection. 

\begin{figure}[htbp] 
   \centering
   \includegraphics[width=6in]{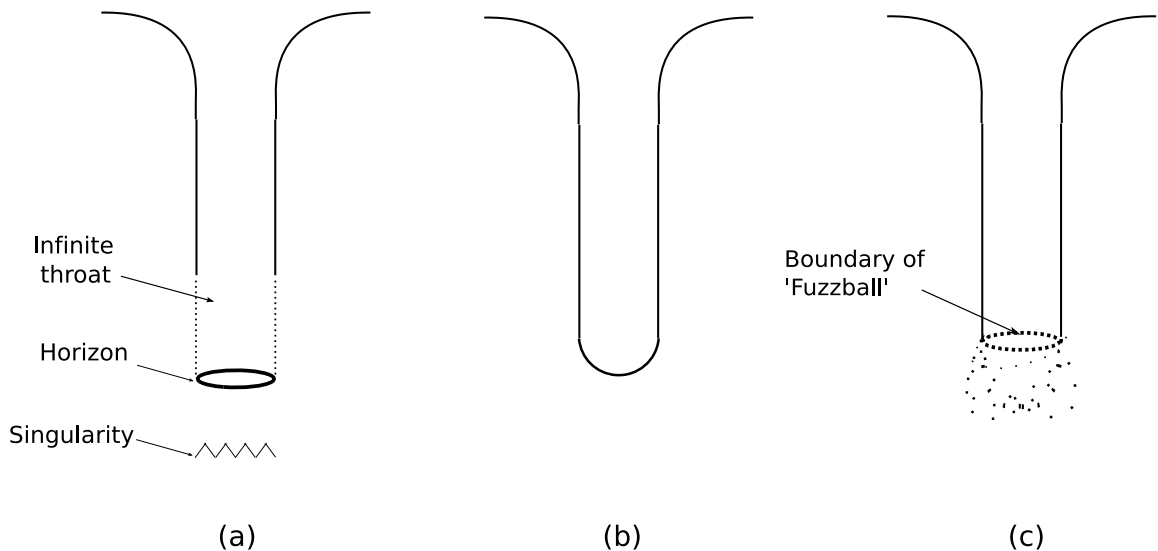} 
   \caption{(a) The `naive geometry of the extremal hole  (b) a special microstate geometry (c) A generic `fuzzball' geometry.}
   \label{first}
\end{figure}

Let us consider extremal holes, using string theory.  Consider first the 3-charge system, which can be made with charges D1-D5-P wrapped on the cycles of a compactification ${\cal M}^4\times S^1$ \cite{sv,cm}. At weak coupling $g_s\r 0$ we can count the microstates that have charges $(n_1, n_5, n_p)$, getting an entropy $S_{micro}$. Next, follow the system towards larger coupling $g_s$. Now we can say that these charges should describe an extremal black hole in 4+1 noncompact dimensions. A classical black hole of this type has an infinitely deep throat, ending in a horizon whose area gives $S_{bek}=S_{micro}$ (fig.\ref{first}(a)).

But because the throat is infinitely deep, the system has no energy gap; we can place a wavepacket in the throat as deep as we want, which means that we can get as much of a redshift as we want. This sounds strange, since we are still dealing with a finite sized system made of a finite number of charges, so we would have expected that  the energy gap to be small but nonzero. 

\b\b

\b\b

In \cite{emission} it was argued that as we increase $g_s$ from zero, the `size' of the brane bound state increases as well, such that at all values of $g_s$ the size of the state is of the same order as the horizon radius at that $g_s$.  The essential reason for this large size was the phenomenon of `fractionation', which makes the brane bound state have a size that depends on the number of charges it contains, instead of being a fixed size like planck length or string length. 

The computation of \cite{emission} was a very crude estimate. But in \cite{lm4} it was noted that one can start with the  2-charge system, which gives the simplest model of a black hole. In this case all states can be understood explicitly in the CFT and in gravity, and the 2-charge bound states were found to be `horizon sized fuzzballs'. Next, we find that we can take very special states of the D1-D5-P system where many quanta are chosen to lie in the same mode. This makes the  state `classical', and we can construct the geometry for these states. One finds the situation in fig.\ref{first}(b), where the throat ends in a smooth cap rather than a horizon \cite{gms}. The energy gap is now finite, and is found to agree exactly with the gap in the dual CFT state. 

The generic state will not have all its quanta placed in a few modes, so we will not get a classical geometry; rather the throat will end in  a  very quantum `fuzzball' (fig.\ref{first}(c)). Instead of the precise energy `gap' mentioned above we can consider a closely related quantity:  the depth of the throat up to the quantum region. An indirect argument shows that this depth again agrees with the corresponding quantity in the CFT \cite{fuzzhigh}.

Large families of 3 charge geometries (in 4+1 dimensions) and 4 charge geometries (in 3+1 dimensions) have now been constructed \cite{bena,bena2,gimon,gimon2,gs}. These all have the same charges and mass as the extremal black hole,  but no horizon; the throat is `capped' in each case.

If the picture fig.\ref{first}(c) is a correct description of the hole, then black holes are not that different from other quantum systems; they are large quantum balls with information distributed throughout a horizon sized region. What basic physics governs the size of these objects?

\subsection{Exploring the role of phase space}

A principal feature of black holes is their large entropy. But entropy is just the log of the volume of phase space occupied by the allowed states of the system. If this phase space is large, perhaps it is not possible to make the system too small in size, and this may be the reason that the generic state becomes a horizon sized `fuzzball'. We will explore this notion  with the help of a small calculation.

\begin{figure}[htbp] 
   \centering
   \includegraphics[width=6in]{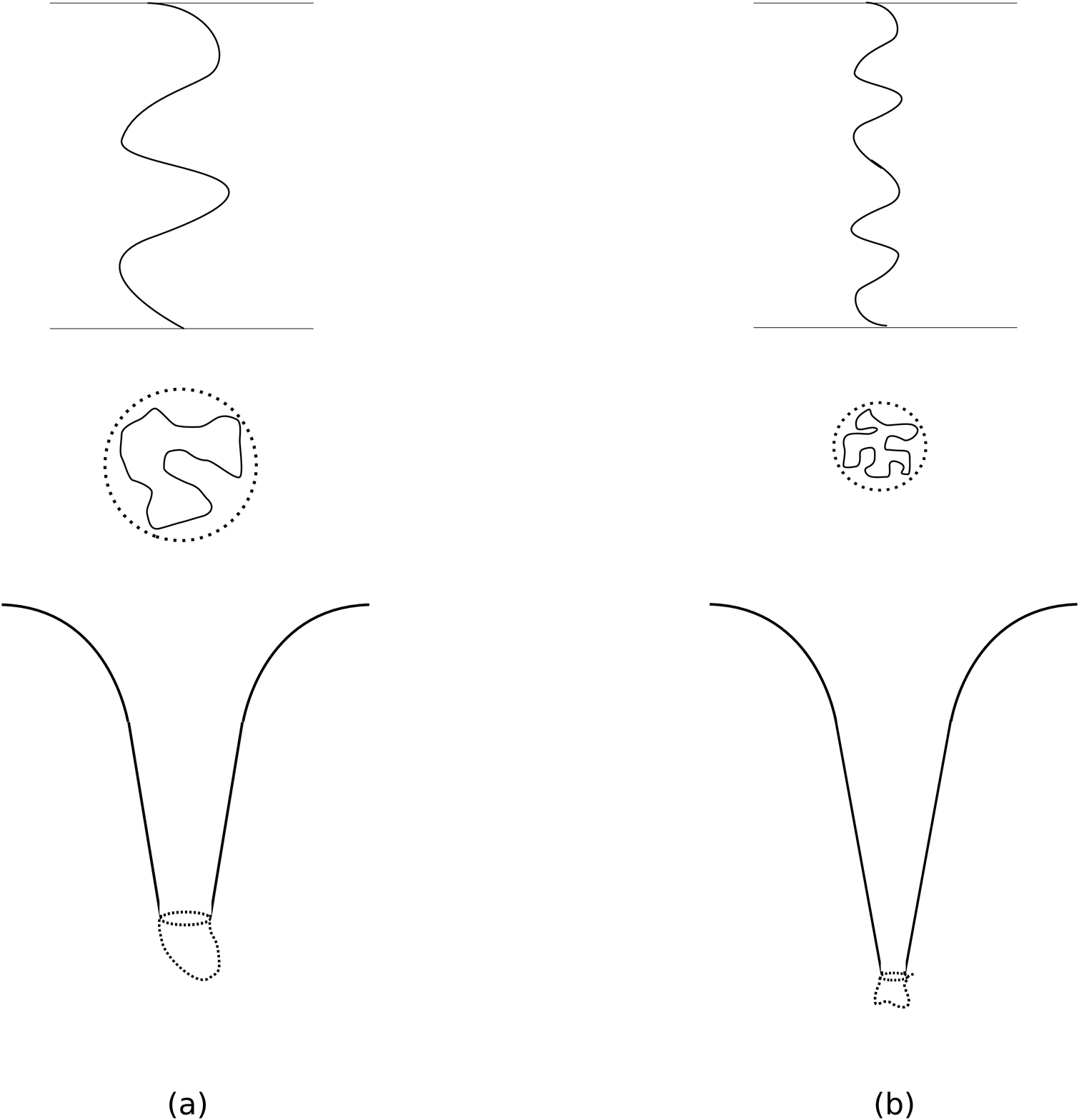} 
   \caption{(a) The first drawing at the top depicts the NS1 string carrying a transverse vibration in a generic state; the string is opened up to its full length. Below that we draw the line traced out by this string in the noncompact directions; this curve lies in a ball shaped region given by the dotted line. Below that we sketch the geometry of the state: typical states of this type differ in the `cap' region drawn by the dotted line, and we wish to find the area of the boundary indicated by the dotted circle.  (b) The same as (a) but for a sub-ensemble of states which have shorter wavelength and smaller transverse displacements.  The `ball' is now smaller, and the throat deeper. We again wish to find the area of the boundary indicated by the dotted circle.}
   \label{intro}
\end{figure}

 Consider the 2-charge system, where we compactify IIB string theory on $T^4\times S^1$. We wrap an NS1 string $n_1$ times along $S^1$, and add $n_p$ units of momentum along $S^1$. The entropy is large since it grows with the charges
\be
S_{micro}=2\sqrt{2}\sqrt{n_1n_p}
\label{en1}
\ee
Now let us picture the states contributing to this entropy. The NS1 is a `multiwound string' wound $n_1$ times around the $S^1$ before joining back to itself. In fig.\ref{intro}(a), at the top, we have drawn this string opened up to its full length, (i.e., we have taken the $n_1$-fold cover of the $S^1$). The momentum is carried by transverse vibrations of the NS1, and  fig.\ref{intro}(a) depicts a `generic state' contributing to the entropy (\ref{en1}), with its corresponding transverse amplitude. Below this string we sketch the trajectory taken by the NS1 in the noncompact directions; we see that the transverse vibrations of the string cause it to spread over a ball in the these noncompact directions. Below this ball we sketch the geometry produced by the string state. It is flat space at infinity, then a throat narrowing towards the location of the string, but this throat does not reach down to a pointlike singularity. Instead,  the throat `caps off' when we reach down to the boundary of the ball. We have taken a classical profile for the NS1, but if we include quantum fluctuations, then the region inside this boundary gives the `fuzzball' region.

It was observed in \cite{lm5} that the area of the boundary of this ball satisfies
\be
{A\over G}\sim \sqrt{n_1n_p}\sim S_{micro}
\label{rel}
\ee
Thus the actual microstates of the system have no horizon, but the boundary of the region where microstates differ from each other satisfies a Bekenstein type relation with the entropy inside the boundary. Can this be an instance of a more general principle? In \cite{lm5} the rotating 2-charge system was also considered; here the microscopic entropy is $2\sqrt{2}\pi\sqrt{n_1n_p-J}$, and boundary area of the fuzzy region around the ring again satisfied the analogue of (\ref{rel}).

To explore further this relation between boundary area and enclosed entropy, we proceed as follows.  We again consider the NS1-P system, but restrict our states to a `sub-ensemble', which we now describe.  The total momentum carried on the string is fixed by the momentum charge that we have chosen. But we can have states where the momentum is in high wavenumbers with small transverse amplitude, or in low wavenumbers with large transverse amplitude. The entropy is optimised for some  typical wavenumber and its corresponding amplitude, and such a typical state has been depicted in  fig.\ref{intro}(a). But now let us restrict to the subset of states which have a wavenumber higher than this optimal value; the amplitude of transverse vibration will then be smaller than the amplitude  in the generic state (fig.\ref{intro}(b)). The microscopic entropy of  states restricted to this smaller amplitude is smaller than (\ref{en1}) by a factor which we call $M$. In the spacetime description, the NS1 now lives inside a smaller `ball', and the spacetime geometry has a throat which becomes narrower and deeper before capping off. We compute the boundary of the fuzzball region in this case, and ask if, to leading order, it has also decreased by a factor $M$. We find that such is indeed the case.
We repeat the computation for the case with rotation, and find a similar agreement. This is the main computation of this paper.

The sub-ensembles with smaller entropy had their fuzzball boundaries at deeper points in the throat, where the transverse area was smaller. We observe that we can recast this computation in a form where we hold the location of the boundary fixed but reduce the {\it distortion} ${\Delta g\over g}$ allowed in the metric at the boundary. Allowing a smaller distortion forces the throat to continue to be the `naive' one to deeper depths, and thus leads to the sub-ensemble with small entropy. 

 We suggest a relation between the `capped' nature of throats  and an old puzzle with extremal holes:  in a Euclidean gravity computation one finds  that the entropy of an extremal hole is {\it zero} \cite{gibbons}. The Euclidean computation uses an infinitely long uniform throat, which does not correspond to a state in the fuzzball picture.   String theory counts actual microstates, and these all `cap off' before reaching the end of the infinite throat.  

Finally, we check that quantum corrections will not change our conclusions. To this end we list the sources of quantum effects, and examine their roles. In the process we compute various length and curvature scales to obtain a picture of the generic 2-charge extremal state for our $T^4\times S^1$ compactification.

We end with some conjectures on how phase space effects may help resolve puzzles associated with black holes.  

\section{The 2-charge system: review}\label{twochargeq}
\setcounter{equation}{0}

Let us recall the basic structure of the 2-charge extremal system. The results summarized below can be found in more detail in the review \cite{fuzzball}.

\subsection{The NS1-P system}

Take type IIB string theory, with the compactification
\be
M_{9,1}\rightarrow M_{4,1}\times {\cal M}_4\times S^1
\ee
The 4-manifold ${\cal M}$ can be $T^4$ or K3; we will let it be $T^4$ for now. Let this $T^4$ have volume $(2\pi)^4 V$, and let the $S^1$ have length $2\pi R$. The string coupling is $g$, and we will set $\alpha'=1$.

Now consider the bound state made with charges NS1-P, where the NS1 charge corresponds to an elementary string wound $n_1$ times on $S^1$, and the P charge is $n_p$ units of momentum along $S^1$. 
The `naive' supergravity solution for these charges is
 \bea
ds^2_{string}&=&H[-dudv+Kdv^2]+\sum_{i=1}^4 dx_idx_i+\sum_{a=1}^4 dz_adz_a\n \\
B_{uv}&=&-{1\over 2}[H-1]\n \\
e^{2\phi}&=&H\n \\
H^{-1}&=&1+{Q_1\over r^2}, ~~\qquad K={Q_p\over r^2}
\label{naive}
\eea
where  $u=t+y, ~v=t-y$, the $z_a$ are coordinates along the $T^4$ and $x_i$ are coordinates in noncompact space. We have
\be
Q_1={g^2\over V} n_1, ~~~~Q_p={g^2\over V R^2} n_p
\label{q1qp}
\ee
As shown in \cite{lm4}, the actual geometries produced by the states of this system are different from the `naive' solution (\ref{naive}). The P charge on the NS1 is carried by transverse vibrations of the NS1. We will focus on the vibrations in the noncompact directions $x_i$ as these will give us an estimate for transverse size of the bound state.\footnote{Vibrations in the $T^4$ were studied in \cite{lmm} and the fermionic excitations were studied in  \cite{marika}.} These vibrations are described by a `profile function' $\vec F(v)$.  The NS1 string, in the bound state, is a single `multiwound' string of total  length
\be
L_T=2\pi R n_1 
\label{tlength}
\ee
The vibration profile is a function of $v=t-y$. 
There are four components of this displacement, $F_1, F_2, F_3, F_4$, collected into the vector $\vec F$.  Thus the noncompact vibrations are described by $\vec F(v)$ with $v$ ranging over $(0, L_T)$. The geometry for a profile $\vec F(v)$ is
 \bea
ds^2_{string}&=&H[-dudv+Kdv^2+2A_i dx_i dv]+\sum_{i=1}^4 dx_idx_i+\sum_{a=1}^4 dz_adz_a\nn
B_{uv}&=&-{1\over 2}[H-1], ~~\qquad B_{vi}=HA_i\nn
e^{2\phi}&=&H
\label{ttsix}
\eea
where
\bea
H^{-1}&=&1+{Q_1\over L_T}\int_0^{L_T}\! {dv\over |\vec x-\vec F(v)|^2}\\
K&=&{Q_1\over
L_T}\int_0^{L_T}\! {dv (\dot
F(v))^2\over |\vec x-\vec F(v)|^2}\\
A_i&=&-{Q_1\over L_T}\int_0^{L_T}\! {dv\dot F_i(v)\over |\vec x-\vec F(v)|^2}
\label{functionsq}
\eea
Note that for $|\vec x|\gg |\vec F(v)|$ we have
\be
K\approx {Q_1\over
L_T}{1\over |\vec x|^2}[\int_0^{L_T}\! {dv (\dot F(v))^2}] 
\ee
so that we identify
\be
Q_p={Q_1\over
L_T}\int_0^{L_T}\! {dv (\dot F(v))^2}\equiv {Q_1}\langle (\dot F(v))^2\rangle
\label{q1p}
\ee

\subsection{The D1-D5 system}

We can perform a sequence of S and T dualities, and map this NS1-P system to the D1-D5 system. Now we have D1 branes wrapped on $S^1$, and D5 branes wrapped on $T^4\times S^1$. We will use `primed' coordinates when we  are in the D1-D5 duality frame. The D1, D5 charges are $n'_1, n'_5$ with
$$n'_1=n_p, ~~~~n'_5=n_1$$
The metric and dilaton are
\be
ds^2_{string}=\sqrt{1\over H_1 H_5}[-(dt'-A_i dx'^i)^2+(dy'+B_i dx'^i)^2]+\sqrt{H_1H_5}dx'_idx'_i+\sqrt{H_1\over H_5}dz'_adz'_a
\label{qsix}
\ee
\be
e^{-2\phi}={H_5\over H_1}
\ee
where the harmonic functions are
\bea
H_5&=&1+{\mu^2Q_1\over  L_T}\int_0^{ L_T} {dv\over |\vec x'-\mu\vec F(v)|^2}\nn
H_1&=&1+{\mu^2Q_1\over
 L_T}\int_0^{ L_T} {dv (\dot
 F(v))^2\over |\vec x'-\mu\vec F(v)|^2},\nonumber\\
A_i&=&-{\mu^2Q_1\over L_T}\int_0^{ L_T} {dv~\dot F_i(v)\over |\vec x'-\mu\vec F(v)|^2}
\label{functionsqq}
\eea
Here $B_i$ is given by
\be
dB=-*_4dA
\label{vone}
\ee
and $*_4$ is the duality operation in the 4-d transverse  space
$x'_1\dots
x'_4$ using the flat metric $dx'_idx'_i$.

In the duality from (\ref{ttsix}) to (\ref{qsix}) we have kept the metric to be unity
at infinity  in the string frame. This has led to a scaling of the  coordinates $\vec x$ 
\be
\vec x' = \mu \vec x
\label{xpx}
\ee
with\footnote{The relations between moduli $g', R', V'$ and the moduli  in the NS1-P frame is given in \cite{fuzzball}.}
\be
\mu={g'\over \sqrt{V'} R'}
\label{mu}
\ee

By contrast the `naive' geometry which one would write for D1-D5 is
\be
ds^2_{naive}={1\over \sqrt{(1+{Q'_1\over r'^2})(1+{Q'_5\over
r'^2})}}[-dt'^2+dy'^2]+\sqrt{(1+{Q'_1\over r'^2})(1+{Q'_5\over
r'^2})}dx'_idx'_i+\sqrt{{1+{Q'_1\over r'^2}\over 1+{Q'_5\over r^2}}}dz'_adz'_a
\label{naived}
\ee
where $r'^2=x'_ix'_i$. Writing $dx'_idx'_i=dr'^2+r'^2d\Omega_3^3$ we see that (\ref{naived}) is the `naive' metric that is conventionally written for the D1-D5 system: it is flat space at infinity, it has a `throat' for $r'^2< Q'_1, Q'_5$, ending in a singularity at $r'=0$. The geometry of the throat is the Poincare patch of $AdS_3\times S^3\times T^4$ (with a periodic identification  $y'=y'+2\pi R'$). In the `actual' geometries (\ref{qsix}) the throat is `capped off'  smoothly at small $r'$, and the information of the configuration is in the shape of the cap.

For later use, we note that 
\be
Q'_1=\mu^2 Q_p={g'\over V'}n'_1, ~~~~~~Q'_5= \mu^2 Q_1={g'} n'_5
\label{q1q5}
\ee

\section{Estimating the size of the generic fuzzball}\label{genericsize}
\setcounter{equation}{0}

While the geometries (\ref{qsix}) have been generated by starting with a classical profile $\vec F(v)$ for the NS1 string, the actual state of the system will correspond to a quantum state of the string and will be a `fuzzball'. We will discuss the various sources of quantum effects in section \ref{nature} below, but for now we are just interested in the location of the `boundary' of this fuzzball; i.e. the surface inside which the typical state begins to differ from the naive geometry (\ref{naived}). In \cite{lm5} it was observed that the area of this boundary  obeyed 
\be
{A\over G}\sim \sqrt{n'_1n'_5}\sim S_{micro}
\ee
In this section we review this result; more details can be found in \cite{fuzzball}. In the next section we will find that a similar result holds for the sub-ensembles that we consider.

\subsection{Nature of the generic state}

Consider the NS1-P system. The large entropy of the system arises from the fact that the momentum P on the NS1 can be partitioned in many ways among different harmonics. Since we are dealing with BPS states, we can look at them at weak coupling and then follow them to strong coupling. At weak coupling we have a string in flat space carrying transverse vibrations. The total energy on the string is
\be
E=P={n_p\over R}={2\pi n_1n_p\over L_T}
\label{tenergy}
\ee
Each excitation of the Fourier mode $k$ carries energy and momentum
\be
e_k=p_k={2\pi k\over L_T}
\ee
Let there be $m^i_k$ excitations of type $i$ in the harmonic $k$ (The index $i$ runs over 8 bosonic excitations and 8 fermionic excitations.)
 Then we must have
\be
\sum_i \sum_k k m^i_k~=~ n_1n_p
\label{equationm}
\ee
 We assign each BPS state the same probability, since it has the same charges and mass (later we will add an additional weighting with angular momentum). The probability distribution is then a thermal one, describing left moving massless quanta in a box of length $L_T$, with temperature
\be
T=\beta^{-1} =[{12 E\over \pi L_T c}]^{1\over 2}={\sqrt{2}\over L_T}\sqrt{n_1n_p}
\label{temperature}
\ee
(Here the central charge $c=12$ describes the effective number of degrees of freedom of the massless gas; The bosons contribute $c=8$ and the fermions $c=4$.) The entropy  is
\be
S_{micro}=2\sqrt{2}\pi\sqrt{n_1n_p}
\label{entropy}
\ee
Consider the bosonic excitations. Since we will only be estimating quantities upto factors of order unity, we restrict to just one flavor of boson and ignore the index $i$. For this boson the occupation number of the $k$th harmonic is
\be
<m_k>={1\over 1-e^{-\beta e_k}}={1\over 1-e^{-{\sqrt{2}\pi\over \sqrt{n_1n_p}}k}}
\label{occupation}
\ee

\subsubsection{Microscopic description of the generic state}

Let us build up a rough picture of how energy is distributed among harmonics in the generic state. Since we just have a thermal gas of excitations, the typical excitation quantum will have energy $\epsilon_k\sim T$, which from (\ref{temperature}) gives for the mean harmonic order
\be
\bar k\sim \sqrt{n_1n_p}
\label{bark}
\ee
From (\ref{occupation}) we see that for modes with $k\sim \bar k$ the occupation number is 
\be
<m_k>\sim 1
\label{unity}
\ee
The number of harmonics that have $k\sim \bar k$  is itself of order $\sim \bar k$. For concreteness let us consider a range
\be
{1\over 2}\bar k < k < {3\over 2}\bar k
\label{range}
\ee
which has a width
\be
\Delta k\sim \bar k \sim \sqrt{n_1n_p}
\ee
The number of quanta in this range is
\be
N\sim \Delta k~ m_k \sim \sqrt{n_1n_p}
\label{nvalue}
\ee
With each quantum carrying energy $\epsilon_{\bar k}$  the total energy is
\be
E\sim  \epsilon_{\bar k} N\sim {2\pi\over L_T}\bar k N\sim {2\pi\over L_T} n_1n_p
\ee
which agrees with (\ref{tenergy}).

\subsubsection{A comment on the definition of `size'}

Consider fig.\ref{intro}(a). In the next subsection we will find the radius of the ball enclosing the curve traced out by the NS1. But there are many subtleties in defining this ball. We will mention some of these here, and then specify how we will define the transverse size of the state. 

In the top sketch of  fig.\ref{intro}(a) we have shown the NS1 carrying transverse oscillations of some typical wavelength $\lambda$. If we had a fourier wave of one precise  wavelength $\lambda$, then the transverse size would be $|\Delta\vec x|\sim |{d\vec F\over dv}|(v)\lambda\equiv |\dot{\vec F}(v)|\lambda$. If we have a spread over fourier modes, and the typical wavenumber is $\bar k$, then in a typical oscillation the string will have a transverse displacement $\sim |\dot{\vec F}(v)|{2\pi\over \bar k}$. The ball in fig.\ref{intro}(a) is drawn with the assumption that  the oscillations of the string keep the string going back and forth in a region with radius of this order, and we will use this picture to compute the ball size below. Thus we are in a sense taking the `smallest possible size' allowed to the ball when the  vibrations have mean wavenumber $\bar k$. 

But there are many other measures possible for the `size' of the state. 
For example, when there is a spread over $k$ we can compute many different `averages', which weight low and high $k$ in different ways. 
In \cite{deboer} a different way of computing `size' was used and a much larger radius for the ball was found. This does not conflict with the essential idea of fuzzballs, since if the ball size is bigger than horizon size the information cannot be trapped inside a horizon. But we would like to see which way of measuring size gives a relation to the area of the boundary, and as found in \cite{lm5} the definition we will use gives a direct relation to the idea of area entropy. 

To see how other definitions of size might be related to the one we will choose, consider again the string in fig.\ref{intro}(a). During each wavelength of oscillation the transverse displacement is $|\Delta\vec x|$. Suppose we say that the path of the string in the noncompact directions is a random walk, with step size $|\Delta\vec x|$. Then we will get a much larger `ball size' than $|\Delta\vec x|$, since the random walk has $N_{steps}\gg 1$  steps and will drift over a large transverse distance. But if we Fourier transform this path, we will see that this large drift corresponds to low harmonics on the string, i.e., modes with $k<<\bar k$. In \cite{lm5} it was noted that the `tail' of low frequency modes in the distribution  makes the entropy from the boundary area larger than the microscopic entropy by a logarithmic factor. We will filter out the effects on size that arise from the tail of low harmonics. But because of these uncertainities in defining the ensemble, {\it we will ignore all log corrections below.}

\subsubsection{Gravity description of the generic state}

Recall that the total length of the NS1 string is $L_T$ given by (\ref{tlength}). Thus for a mode in harmonic $\bar k$ the wavelength of vibrations is
\be
\lambda={L_T\over \bar k}\sim {2\pi R n_1\over \sqrt{n_1n_p}}\sim R\sqrt{n_1\over n_p}
\label{wavelength}
\ee
We wish to ask how much the transverse coordinate $\vec x$ changes in the process of oscillation. Thus we set $\Delta y=\lambda$, and find
\be
|\Delta\vec x|\sim |\dot{\vec F}|\Delta y\sim |\dot{\vec F}|R\sqrt{n_1\over n_p}
\label{step1}
\ee
From (\ref{q1p}) we know that
\be
Q_p \sim Q_1 |\dot{\vec F}|^2
\ee
which gives
\be
| \Delta\vec x|\sim \sqrt{Q_p\over Q_1}R\sqrt{n_1\over n_p}\sim 1
\label{ffone}
\ee
where we have used (\ref{q1qp}). 

We wish to compute the `size' of a generic fuzzball state. It turns out that it is easier to understand quantum corrections in the D1-D5 frame than in the NS1-P frame \cite{fuzzhigh}. Thus we will work in the D1-D5 frame below.  In this duality frame the `spread' (\ref{ffone}) of the state in coordinate terms is
\be
r'\equiv |\Delta \vec x'|=\mu|\Delta\vec x|\sim \mu ={g'\over \sqrt{V'} R'}
\label{bvalue}
\ee
where we have used (\ref{mu}).

For 
\be
|\vec x'|\gg |\Delta\vec x'|
\ee
we have
\be
{1\over |\vec x'-\mu\vec F|^2}\approx {1\over |\vec x'|^2}
\ee
and the solution becomes approximately the naive geometry (\ref{naived}).

We see that the metric settles down to the naive metric outside a certain ball shaped region $|\vec x'|>\mu$. 
Let us now ask the question: What is the surface area of this ball?
To compute this, we will take the naive metric (\ref{naived}), put a surface at $r'=|\vec x'|=\mu$, and find the area of this surface. Since we will need a more general result in a moment, let us set 
\be
r'=b
\ee
and compute the area of the bounding surface in the naive geometry (\ref{naived}). In the Bekenstein relation we can compute the area of the horizon surface in the 10-d Einstein metric and divide by $G_N^{(10)}$, or compute the area in the 5-d Einstein metric and divide by $G_N^{(5)}$
\be
{A_E^{(10)}\over G_N^{(10)}}= {A_E^{(5)}\over G_N^{(5)}} 
\label{newton}
\ee
We will compute the area in 10-d, where we have  an 8-dimensional surface: an $S^1$ from the direction $y'$, a $T^4$ from the directions $z'_a$, and an $S^3$ from the angular directions in $\vec x'$. We first compute in the string frame, using the metric (\ref{qsix}), and then convert to the Einstein frame. 

In the string frame the $S^1$ has a length
\be
L_y=(2\pi R') (H_1H_5)^{-{1\over 4}}
\label{length1}
\ee
The angular sphere in the $\vec x'$ directions gives
\be
V_{S^3}=2\pi^2 b^3 (H_1H_5)^{3\over 4}
\ee
The $T^4$ gives
\be
V_4=(2\pi)^4 V' ({H_1\over H_5})
\ee
Converting to the Einstein metric $g_E$ from the string metric $g_S$ is done through
\be
g_E= e^{-{\phi\over 2}}g_S=({H_1\over H_5})^{-{1\over 4}}g_S 
\label{convert}
\ee
With all this we find the area of our surface in the Einstein metric
\be
A_E^{(10)}={1\over 2} (2\pi)^7 (H_1H_5)^{{1\over 2}} b^3 V' R'
\ee
Writing
\be
H_1\approx {Q_1\over b^2}, ~~~H_5\approx {Q_5\over b^2}
\ee
we get
\be
A_E^{(10)}={1\over 2} (2\pi)^7 (Q_1Q_5)^{{1\over 2}}  V' R'~b
\ee
Finally, using 
\be
G_N^{(10)}=8\pi^6 g'^2
\ee
we find
\be
{A_E^{(10)}\over G_N^{(10)}}\sim (Q_1Q_5)^{{1\over 2}} { V' R'\over g'^2} ~b
\label{areafinal}
\ee
where we have now dropped all numerical constants since we can specify the value $b\sim \mu$ only upto  a factor of order unity. Setting $b$ to the value (\ref{bvalue}) we get
\be
{A_E^{(10)}\over G_N^{(10)}}\sim (Q_1Q_5)^{{1\over 2}} { V' R'\over g'^2} ({g'\over \sqrt{V'} R'})\sim \sqrt{n'_1n'_5}
\ee
where we have used (\ref{q1q5}).

Noting (\ref{entropy}), we now observe that \cite{lm5}
\be
{A_E^{(10)}\over G_N^{(10)}}\sim S_{micro}
\label{relation}
\ee
Thus we see that the size of the `fuzzball' made by the generic state is such that its surface area satisfies a Bekenstein type relation with the entropy of the fuzzball.\footnote{For the heterotic NS1-P system  a relation like (\ref{relation}) was derived in \cite{sen} using arguments based on a `stretched horizon'.} It was argued in \cite{lm5} that if a quantum falls into the region enclosed by this boundary, the complicated nature of the interior state traps the quantum for very long times, so that this boundary behaves like a `horizon' for observers that are observing the system only over classical timescales: i.e. times of order the radius of  the $S^3$. 

\section{Boundary area for sub-ensembles}\label{sub}
\setcounter{equation}{0}

In this section we will perform the main computation of this paper. We wish to get some understanding of why the relation (\ref{relation}) turned out to be true. As discussed in the introduction, black holes have a large entropy, and entropy is a measure of phase space volume. One might therefore conjecture that the `fuzzball' cannot be too small because it needs to contain enough phase space to account for this large entropy. 

Note that the area $A_E^{(10)}$ in (\ref{relation}) is `large' in terms of the elementary scales in the theory. For the D1-D5 system, the dilaton approaches a constant down the `throat' of the geometry, so  string length $l_s$ and the 10-d planck length $l_p$ are of the same order everywhere. In other words, 
the ratio $l_s/l_p$ does not depend on the charges, which we will take to satisfy $n'_1, n'_5\gg 1$. Since (\ref{relation})
says that the horizon area in planck units gives the entropy $\sim \sqrt{n'_1n'_5}$,  we see the 10-d horizon area is much larger than planck or string scale. From (\ref{newton}) we see that the horizon area is large in terms of the 5-d planck scale as well.

\subsection{The sub-ensemble}

How do we explore the role of phase space in determining the size of the fuzzball? As explained in the introduction, we will take a subset of   states  -- which we call a `sub-ensemble' -- which contains states with transverse spread smaller than the spread of the generic state. We can compute the area of the bounding surface for the typical state of this sub-ensemble; this will be smaller than the area $A_E^{(10)}$ in (\ref{relation}) by some factor $1/M$ (with $M>1$).  On the other hand, the entropy of such states will also be smaller than $S_{micro}$ by some factor, and we wish to check if this factor is also $1/M$. We will find that the factors indeed agree (up to log terms that we will ignore). In the next section we will repeat the analysis for the 2-charge extremal `black ring', which is the same 2-charge system but with rotation. We will again find a similar agreement. 

\subsubsection{Microscopic description of the sub-ensemble}\label{mic}

Let us discuss the microscopic description first. As mentioned above, we will be working with the D1-D5 duality frame later, so 
we will make the replacement
\be
n_1 ~\r ~ n'_5, ~~~~~n_p~\r ~ n'_1
\ee
in all relations which had been developed in the NS1-P duality frame.

Consider the entropy  (\ref{entropy}). We want to get a crude argument for why this entropy is order $\sqrt{n'_1n'_5}$ in the charges. From (\ref{bark})  we see that the generic quantum is in harmonic $k\sim\sqrt{n'_1n'_5}$, and from (\ref{nvalue}) we note that there are
$N\sim \sqrt{n'_1n'_5}$ such quanta. From (\ref{unity}) we have that $\langle m_k\rangle \sim 1$ for these modes $k$. Let us  thus take the crude approximation that this occupation number can be $0$ or $1$ for each mode in the range (\ref{range}), with equal probability. The number of possibilities is of order
\be
{\cal N}\sim 2^{\sqrt{n'_1n'_5}}
\ee
so that the entropy is
\be
S_{micro}\equiv \ln {\cal N} \sim \sqrt{n'_1n'_5}
\ee
We could certainly make a much better estimate if we wished; for  example we can take into account the fact that each mode has several polarization states, or the fact that the occupation numbers are not just $0,1$, or the fact that occupation probability is not uniform in the range 
(\ref{range}). But the above crude estimate is enough for our purposes, since we will extend it to the case that we now wish to cover.

We are interested in looking at the subset of states where the typical harmonic order is not $\bar k\sim \sqrt{n'_1n'_5}$ but rather
\be
\bar k_M\sim M{\sqrt{n'_1n'_5}}
\label{knew}
\ee
for some $M>>1$. Since the total energy is still (\ref{tenergy}) the number of excitation with $k$ of this order will be
\be
N_M\sim {\sqrt{n'_1n'_5}\over M}
\label{excitations}
\ee
As before, we interpret (\ref{knew})  as saying that $k$ lies in a range
\be
{1\over 2}M{\sqrt{n'_1n'_5}} < k < {3\over 2} M{\sqrt{n'_1n'_5}}
\label{rangenew}
\ee
so that the range of $k$ has a width
\be
\Delta k_M \sim M{\sqrt{n'_1n'_5}} 
\ee
Note however that this time the number  of excitations $N$ in (\ref{excitations}) is much less than the number of elements in the range (\ref{rangenew}) 
\be
{N_M \over \Delta k_M}\sim {1\over M^2} << 1
\label{less}
\ee
So our task is to count the number of ways that $N_M$ excitations can be distributed in the range $\Delta k_M$.
Because of (\ref{less}), we see that the $N_M$ excitations will be sparsely distributed in the range $\Delta k_M$,  so we do not have to worry about more than one excitation falling on the same value of $k$. Each excitation can be in one of  $\Delta k_M$ possible values of $k$, giving $(\Delta k_M)^{N_M}$, and we must then divide by $N_M!$ since we can permute the $N_M$ excitations without changing the state. We thus get for the number of possibilities
\be
{\cal N_M}\sim {1\over N_M!}(\Delta k_M)^{N_M}
\ee
Thus the entropy is
\bea
S_{micro, M}&=&\ln {\cal N_M} ~\approx ~ N_M\ln {\Delta k_M} - [ N_M\ln N_M-N_M]\nn
&=& N_M~ [\ln {\Delta k_M\over N_M}+1]\nn
&\sim& N_M~ [ \ln {M^2}+1] \nn
&\sim & {\sqrt{n'_1 n'_5}\over M}~ [ \ln {M^2}+1]
\eea
where in the third line we have used (\ref{less}) and in the last line we have used (\ref{excitations}).
As mentioned above, we will drop all log corrections; we will just note that the power of $M$ by which $S_{micro, M}$ changes  is the power $M^{-1}$.    So we just write
\be
S_{micro,M}\sim {\sqrt{n'_1n'_5}\over M}
\label{mfinal}
\ee

\subsubsection{Gravity description of the sub-ensemble}

Let us now turn to the gravity description of the states in our sub-ensemble. 
Instead of the wavelength $\lambda$ in (\ref{wavelength}) we have the wavelength
\be
\lambda_M={L_T\over \bar k_M}\sim {1\over M} R\sqrt{n_1\over n_p}
\label{wavelengthp}
\ee
Following the same steps as in eq.(\ref{step1})-eq.(\ref{ffone}) we get (in place of (\ref{ffone}))
\be
|\Delta \vec x|\sim  {1\over M}
\label{ffonep}
\ee
Thus the  corresponding coordinate size in the D1-D5 frame is (using (\ref{xpx}))
\be
|\Delta \vec x'|={1\over M}{g'\over \sqrt{V'} R'}
\label{bvaluenew}
\ee
We must therefore set
\be
b={1\over M}{g'\over \sqrt{V'} R'}
\ee
in (\ref{areafinal}). Calling the area of the boundary $A_{E, M}^{(10)}$ we get
\be
{A_{E, M}^{(10)}\over G_N^{(10)}}\sim (Q_1Q_5)^{{1\over 2}} { V' R'\over g'^2} ({1\over M}{g'\over \sqrt{V'} R'})\sim {1\over M}\sqrt{n'_1n'_5}
\ee
Comparing with (\ref{mfinal}) we see that 
\be
{A_{E, M}^{(10)}\over G_N^{(10)}}\sim S_{micro, M}
\ee
This is the main computation of this paper. We will next see that a similar result of obtained for the system with rotation. We will comment on  physics of these results in section \ref{what} below.  

\section{The 2-charge black ring}\label{ringsec}
\setcounter{equation}{0}

We can give our 2-charge extremal system an angular momentum, and obtain a 2-charge {\it black ring}. In \cite{lm5} this system was also considered, and it was again found that the area $A$ of the boundary of the fuzzy region   gave ${A\over G}\sim S_{micro}$.  We would  like to repeat for this system the steps of the previous section; i.e., we restrict to a  sub-ensemble with smaller entropy, and then look for the area of the bounding surface for this sub-ensemble. We will again find an agreement similar to the one for the non-rotating system.

\b\b

\begin{figure}[htbp] 
   \centering
   \includegraphics[width=5in]{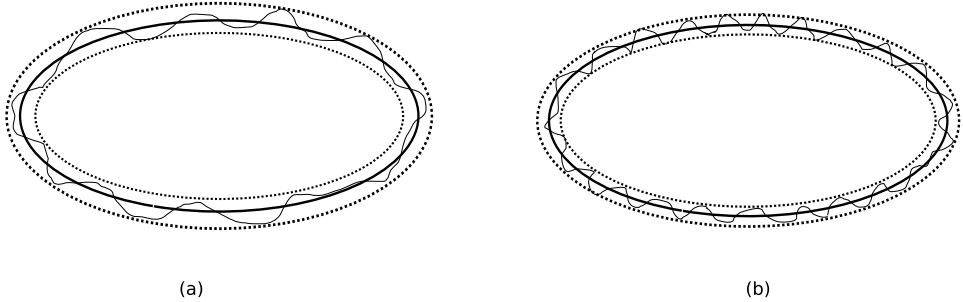} 
   \caption{(a) In the generic state with angular momentum $J$ the NS1 describes a path close to a circle; the tube shaped boundary encloses the region inside which the typical curve fluctuates  (b) The paths for the sub-ensemble. The vibrations have smaller amplitude and higher wavenumber, and  the tube is therefore `thinner' with a smaller surface area.}
   \label{ring}
\end{figure}

\b\b

\subsection{The full ensemble: review}

We begin by recalling the computation of \cite{lm5}. 
Let us start with  the NS1-P frame. The charges are still $n_1, n_p$, but we require the system to have an angular momentum $J$. In \cite{lm3} it was shown that $J\le n_1n_p$, so we will set 
\be
J=\alpha n_1n_p=\alpha n'_1n'_5, ~~~~0<\alpha\le 1
\ee
To give the system an angular momentum in the $x_1-x_2$ plane, we let the NS1 string have a profile
\be
{\vec F}(v)=a~{\vec e}_1\cos\frac{2\pi v}{ L_T}+a~{\vec e}_2\sin\frac{2\pi
v}{ L_T}+{\vec X}(v)
\label{residual}
\ee
The path traced out by the NS1 now lies close to a circle in the $x_1-x_2$ plane; this circle has radius
$a=|\vec x|=\sqrt{J}$. Since we will be working with the D1-D5 system, we note the radius of this circle in the coordinates $\vec x'$
\be
a'=\mu \sqrt{J}=\frac{g'}{\sqrt{V'}R'}\sqrt{J}
\ee

\subsubsection{The microscopic computation}

Consider the NS1-P state. We  again have to put momentum  excitations on the NS1 satisfying (\ref{tenergy}), but with the constraint that the state carries an angular momentum $J$. The angular momentum can be carried by the bosonic excitations as well as the fermionic excitations. But if  $J$ is large enough to be  of order $\sim n_1n_p$, we find that there is a simplification. A large number of bosonic excitations drop to the lowest harmonic $k=1$ and carry all the angular momentum $J$; these modes therefore contribute nothing to the entropy. The rest of excitations distribute themselves among all harmonics $k$ to yield the entropy, but they are not constrained by the requirement of carrying any net angular momentum. The $J$ excitations carrying the angular momentum have energy 
\be
E^{rotation}={2\pi\over L_T}J
\label{erot}
\ee
so the remaining modes carry energy
\be
E^{vibrations}={2\pi\over L_T}( n_1n_p-J)
\label{evib}
\ee
The entropy is therefore \cite{lm5,rings}
\be
 S_{micro}=2\sqrt{2}\pi\sqrt{n_1n_p-J}=2\sqrt{2}\pi\sqrt{n'_1n'_5-J}
\label{ringm}
\ee

Let us also note the nature of the generic state.  The energy (\ref{evib}) has to be distributed between different harmonics. Thus we will have, in place of (\ref{bark}),  a mean harmonic order
\be
\bar k\sim \sqrt{n'_1n'_5-J}
\label{kj}
\ee
The number of quanta with wavenumber of this order will be (in place of (\ref{nvalue}))
\be
N\sim \sqrt{n'_1n'_5-J}
\ee

\subsubsection{The boundary of the `fuzzy ring'}
 
 Consider the path of the NS1 in the $\vec x$ space. We depict this in fig.\ref{ring}(a). The central circle is the circle of radius $a$, and the residual vibrations $\vec X(v)$ in (\ref{residual}) make the actual NS1 profile wiggle around in a thin tube shape region around this circle. We draw a tube shaped boundary  to bound the region occupied by the NS1 in a typical state. The area of this boundary (in the $\vec x'$ plane) will therefore be given by the area of an $S^2$ transverse to the circle, times the length of an $S^1$ along the circle direction. 
 
We must first find the coordinate radius of this $S^2$ which measures the typical displacement of the NS1 from its mean location on the circle. This takes a little  more work than in the non-rotating case. With $\bar k$ given by (\ref{kj}) the analog of (\ref{wavelength}) is 
\be
\lambda={L_T\over \bar k}\sim { R n_1\over \sqrt{n_1n_p-J}}
\label{wavelengthj}
\ee
Let us go to a particular point on the ring. In the vicinity of this point, the vector $\vec F(v)$ can be decomposed into two parts
\be
\vec F(v)=\vec F_{rot}(v)+\vec F_{vib}(v)
\ee
Here $\vec F_{rot}$ is the part that comes from the first two terms on the RHS of (\ref{residual}); thus it points along the ring and gives it its angular momentum $J$. The part $\vec F_{vib}$ comes from $\vec X(v)$ in (\ref{residual}). We have
\be
|\dot{\vec F}(v)|^2=|\dot{\vec F}_{rot}(v)|^2+|\dot{\vec F}_{vib}(v)|^2+2 \dot{\vec F}_{rot}(v)\cdot\dot{\vec F}_{vib}(v)
\ee
We will assume that the randomness of the fluctuations $\vec X(v)$ gives
\be
\dot{\vec F}_{rot}(v)\cdot\dot{\vec F}_{vib}(v)=0
\ee
We have
\be
|\dot{\vec F}_{rot}(v)|^2={(2\pi a)^2\over L_T^2}=({2\pi\over L_T})^2 J
\ee
The relation (\ref{q1p}) then gives that
\be
Q_p=Q_1[({2\pi\over L_T})^2 J+\langle |\dot{\vec F}_{vib}(v)|^2\rangle]
\ee
Thus
\be
|\dot{\vec F}_{vib}(v)|=\sqrt{ {Q_p\over Q_1}-({2\pi\over L_T})^2 J}={1\over n_1 R}\sqrt{n_1n_p-J}
\ee
where we have used (\ref{q1qp}).
The transverse displacement of the profile from the central circle of the ring is therefore given by a computation analogous to (\ref{step1})
\be
|\Delta \vec x|\sim |\dot{\vec F}_{vib}(v)|\lambda\sim 1
\ee
where we have used (\ref{wavelengthj}).
Thus the radius $r'_{S^2}$ of the sphere transverse to the ring is of the same order  as the transverse displacements (\ref{ffone}) in the non-rotating case. Transforming to the D1-D5 frame, we have
\be
|\Delta \vec x'|=r'_{S^2} \sim {g'\over \sqrt{V'} R'}
\label{rfinal}
\ee

\subsubsection{Computing the area of the  boundary}
 
As in the non-rotating case, we switch to the D1-D5 frame for our computations.  
 We will again compute the area in 10-dimensions, and divide by $G_N^{(10)}$. Outside the tube shaped boundary, the geometry will be close to a  `naive' metric. We construct this naive metric by taking the general metric (\ref{qsix}), and replacing
$ |\vec x'-\mu \vec F(v)| $ by the distance of $\vec x'$ from the central circle $|\vec x'|=a'$.  We find that the functions in (\ref{qsix}) are given by
\bea
H_1&=&1+\frac{Q'_1}{f_0},~~~~~~~~~~ H_5=1+\frac{Q'_5}{f_0},\nn
A_i dx^i&=&\sqrt{\frac{J}{n'_1n'_5}}
\frac{2\sqrt{Q'_1Q'_5}a'}{f_0({\vec x'}\cdot {\vec x'}+a'^2+f_0)}
(x'_2dx'_1-x'_1dx'_2),\nonumber\\
f_0&=&\left[({\vec x'}\cdot {\vec x'})^2+2a'^2(x'^2_3+x'^2_4-x'^2_1-x'^2_2)+
a'^4\right]^{1/2}
\label{bigeq}
\eea
For $\vec x'$ far outside the tube shaped region in fig.\ref{ring}(a), the actual metric will agree closely with this naive metric. 
 
The length of the $S^1$ in the direction $y'$  is (in the string metric)
\be
L_y=(2\pi R') (H_1H_5)^{-{1\over 4}}
\ee
The $T^4$ gives
\be
V_4=(2\pi)^4 V' ({H_1\over H_5})
\ee
In the noncompact space $\vec x'$, the boundary of the tube has an $S^2$ transverse to the ring and an $S^1$ along the ring. The $S^2$ has area
\be
A_{S^2}=4\pi (r'_{S^2})^2(H_1H_5)^{1\over 2}
\ee
Let $z$ be a coordinate along the ring, normalized so that $|dz|=\sqrt{dx'_idx'_i}$.  Close to the ring, the field $A_i$ has only a component along $z$. From (\ref{bigeq}) we find
\be
A_z=-\sqrt{J\over n'_1n'_5}{\sqrt{Q'_1Q'_5}\over 2a'r'_{S^2}}
\ee
The metric along the ring is then
\be
g_{zz}=\sqrt{H_1H_5}-{1\over \sqrt{H_1H_5}} A_z^2={1\over \sqrt{H_1H_5}}{Q'_1Q'_5\over (2a'r'_{S^2})^2 n'_1n'_5} (n'_1n'_5-J)
\ee
Thus the circle $S^1$ along the ring has length
\be
L_1=2\pi a' \sqrt{g_{zz}}=(H_1H_5)^{-{1\over 4}}{\sqrt{Q'_1Q'_5\over n'_1n'_5}}{\pi\over r'_{S^2}}\sqrt{n'_1n'_5-J}
\ee
Taking the product of the contributions from $L_{y'}, V_4, A_{S^2}, L_1$, and converting to the Einstein metric using (\ref{convert}), we arrive at
\be
A_E^{(10)}=(2\pi)^7\sqrt{Q'_1Q'_5\over n'_1n'_5} \sqrt{n'_1n'_5-J} ~V' R' ~r'_{S^2}
\label{areaj}
\ee
Using (\ref{rfinal}), we get \cite{lm5}
\be
{A_E^{(10)}\over G_N^{(10)}}\sim \sqrt{n'_1n'_5-J}\sim S_{micro}
\ee
where at the last step we have compared with (\ref{ringm}).

\subsection{The sub-ensembles for the black ring}

We will see that all computations for this case are very similar to those in the non-rotating case.

\subsubsection{Microscopic description of the sub-ensemble}

We have to hold fixed the conserved quantities of the ring -- the charges $n'_1, n'_5$ and the angular momentum $J$ -- and then look for a subset of states that define our sub-ensemble. Thus we will look at states where we still have the energy (\ref{erot}) from the lowest harmonic carrying the angular momentum $J$, but the energy (\ref{evib}) from the remaining excitations are taken to be in harmonics of order
\be
\bar k_M \sim M{\sqrt{n'_1n'_5-J}}
\label{bkj}
\ee
with $M\gg 1$. Since the energy $E^{vib}$ is unchanged, the number of excitations with $k$ of this order will be 
\be
N_M\sim {\sqrt{n'_1n'_5-J}\over M}
\label{excitationsj}
\ee
Following the steps in section (\ref{mic}) we find
\be
S_{micro}\sim {\sqrt{n'_1n'_5-J}\over M}
\label{mfinalj}
\ee

\subsubsection{Gravity description of the sub-ensemble}

Fig.\ref{ring}(b) shows the nature of our sub-ensemble. By letting $E^{vib}$ be carried by harmonics $k\sim M\sqrt{n'_1n'_5-J}$ instead of $k\sim \sqrt{n'_1n'_5-J}$ we have put the energy in vibrations that have smaller amplitude and higher wavenumber. Because of the smaller amplitude the vibrations stay closer to the central circle of the ring. Thus the $S^2$ surrounding the ring will have a smaller area, while the length of the direction along the ring will remain unchanged. 

Putting  $\bar k_M$  ((eq.\ref{bkj}))  in  (\ref{wavelengthj}) instead of $\bar k$, and following the steps to  (\ref{rfinal}) we find
\be
r'_{S^2} \sim {1\over M}{g'\over \sqrt{V'} R'}
\label{rfinalj}
\ee
Putting this value of $r'_{S^2}$ in (\ref{areaj}) we get
\be
{A_E^{(10)}\over G_N^{(10)}}\sim {1\over M}\sqrt{n'_1n'_5-J}\sim S_{micro}
\ee
where at the last step we have compared with (\ref{mfinalj}).

\section{What is the role of phase space in the geometry of the fuzzball?}\label{what}
\setcounter{equation}{0}

Black holes have an entropy given by the area $A$ of their horizon. 
But this fact does not tell us {\it why} the area is a measure of the entropy. We cannot answer this question fully at this stage. But in the above sections we have tried to probe the relation between entropy and area more deeply, by looking at sub-ensembles of the full space of states. Let us first summarize the physics that we have seen.

\begin{figure}[htbp] 
   \centering
   \includegraphics[width=6in]{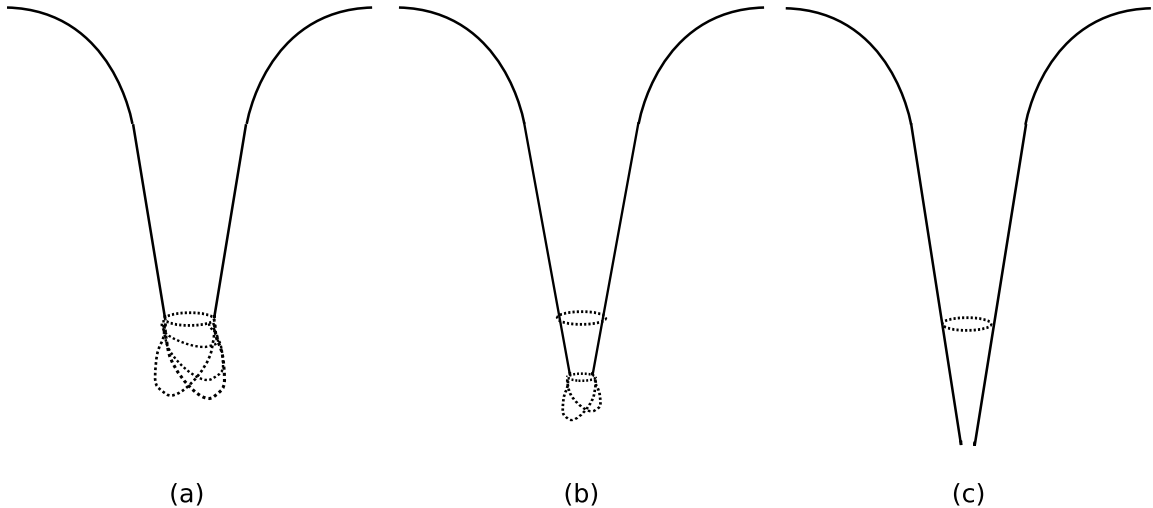} 
   \caption{(a) The 2-charge geometry with the full ensemble of `caps'  (b) a sub-ensemble, which has deeper throats, and therefore a smaller boundary area for  the fuzzball region (c) The naive geometry is taken to continue deeper than the deepest allowed cap. There are no states possible at the end of this throat.}
   \label{twocharge}
\end{figure}

Consider the 2-charge system in fig.\ref{twocharge}. In 
\ref{twocharge}(a), the dashed boundary is placed at the location of the generic `cap'. The size of the $S^1$ at the location of this boundary is estimated by setting $r'=b={g'\over \sqrt{V'} R'}$ (eq.  (\ref{bvalue})) in the naive geometry (\ref{naived}). Since this boundary captures the generic state, the entropy of allowed states inside this boundary is
$S\sim \sqrt{n'_1n'_5}$. 

 Fig.\ref{twocharge}(b) depicts the situation with the sub-ensembles that we considered. The naive geometry goes upto a deeper point, where the $S^1$ is smaller, and the boundary of the region where the caps differ is therefore smaller than the area in fig.\ref{twocharge}(a) by some factor $M$.  The number of `caps' that come after this depth of throat is smaller too, with an entropy $S\sim {1\over M} \sqrt{n'_1n'_5}$. 

In fig.\ref{twocharge}(c) we have let the geometry be the naive one upto a distance longer than the longest throat allowed for any microstate. The $S^1$ boundary is now even smaller, and with this value of the boundary parameters there are no `capped' states possible.


What physics is suggested by these observations? The fuzzball conjecture says that black holes are just large quantum balls, not fundamentally different from any other quantum system. Thus different microstates of the fuzzball must be orthogonal wavefunctions of the system. One can imagine that it is not easy to fit too many orthogonal wavefunctions in a given region (with a given energy). The number of orthogonal states reflects the phase space volume occupied by the ensemble. It is plausible that a large phase space volume implies, for the stationary states of our system,  a large size in ordinary space $\vec x$. In that case the large size of fuzzballs would be the result of the large phase space of states they describe. 

  But if this reasoning were correct, then we might expect a more general relation, which would say that {\it any} sub-ensemble of states with entropy $S_{micro}$ would have to have a boundary area that satisfied  $  {A\over G}\gtrsim S_{micro}$. This is what we found for our sub-ensembles; in fact they happened to be `optimally packed' sets of states, since they achieved the maximal entropy  allowed for their boundary area.\footnote{In \cite{denefmoore} it was argued that there are large numbers of BPS multi-black hole bound states. At large coupling these are expected to be loosely bound, thus giving systems  with very large transverse size. Our present interest is in a somewhat different question. We wish to argue that we cannot fit more states in a given region than allowed by the Bekenstein bound; the existence of other, much larger  sized   states with the same mass and charges is therefore not in conflict with our goal.}
  
It would be interesting to check how these ideas work for 3-charge and 4-charge systems. At present it is not clear to us how sub-ensembles should be made for this case.  

\subsection{Another way to define sub-ensembles}

Suppose we take the sub-ensemble in fig.\ref{twocharge}(b) but place the boundary at the same location as the boundary in 
 fig.\ref{twocharge}(a) (shown by the upper ring in \ref{twocharge}(b)). What will we see at this boundary? For the states in the full ensemble in  fig.\ref{twocharge}(a), the geometry at the boundary $r'={g'\over \sqrt{V'} R'}$ is distorted by order unity compared to  the naive geometry, since the caps start to differ significantly from each other at this point.  For the states in the sub-ensemble of fig.\ref{twocharge}(b), there is very little distortion at this location; the distortions will come much further down the throat at $r'={1\over M} {g'\over \sqrt{V'} R'}$. 
If we place the boundary at $r'={g'\over \sqrt{V'} R'}$  in the geometry of fig.\ref{twocharge}(c) then we will see virtually {\it no} distortion.

Thus we see that there are two geometric quantities that we can consider at a boundary. One is the area $A$ of this boundary. The other is the width ${\Delta g\over g}$ of distortions that we allow at this boundary. We can now describe the ensembles in  
fig.\ref{twocharge}(a),(b),(c) in two equivalent ways. In the first, we place the boundary at the location where the distortions become order unity, and then use the area $A$ at this boundary to differentiate between the ensembles. In the second we keep the boundary at any fixed location, say $r'={g'\over \sqrt{V'} R'}$, and change the allowed width ${\Delta g\over g}$. For the case in fig.\ref{twocharge} (a) we will have a width of order unity, in \ref{twocharge}(b) the width would be smaller, and in \ref{twocharge}(c) it would be virtually zero. 

As we will see in the next subsection, the second way of defining sub-ensembles may be related to other approaches to entropy like the use of asymptotic symmetries.

\subsection{Possible relations to other approaches to entropy}

The principal feature that we see in fuzzballs is that each microstate has a different structure after some distance down the throat. Thus we see the `hair' responsible for the entropy of the system. With this explicit picture of the `hair' we might be able to understand better some other approaches to entropy:

\b

(i) The distortions at the fuzzball boundary give an {\it entanglement} \cite{sred}  
 between the states inside the cap and the geometry just outside. In the fuzzball picture, the latter geometry is not the featureless naive geometry outside a horizon, but reflects the distortions of the `cap'. If we allow virtually no distortions, as in fig.\ref{first}(a), we get no states.

\b

(ii) There is evidence of  {\it holography} \cite{susskind}, since the entropy in the possible `caps' is controlled by the area $A$ and width ${\Delta g\over g}$ of fluctuations at the {\it boundary}. In a general statistical system there would be no reason for the boundary data to reflect the entropy in the interior.

\b

(iii) It may be possible to relate the  ${\Delta g\over g}$ caused by the fuzzball caps to the algebra of diffeomorphisms studied in \cite{brown}. The central charge and level of this algebra appear to give a count of states, but it has never been explicitly clear what is being counted. 
Black hole entropy has also been derived as a Noether charge for diffeomorphisms (in the time direction) \cite{wald}, but again it is not clear what this charge is counting. The fuzzball caps naturally generate distortions, and holography suggests that all information of the cap is captured by the distortions at the boundary.  Thus any method of counting these distortions at the boundary should relate the  entropy to the boundary properties.

\section{An old puzzle with the extremal hole} \label{puzzle}
\setcounter{equation}{0}

The entropy of Schwarzscild holes is given by the area of their horizon. This is also true for holes with charge, but what about {\it extremal} holes, which have the maximum charge for their mass? The extremal Reissner-Nordstrom hole also has a nonvanising horizon area $A$, but it has been argued that its entropy should be {\it zero} \cite{gibbons}. In this argument one notes that the extremal hole has an infinitely long throat and therefore a different topology from nonextremal holes. For nonextremal holes one must compactify the Euclidean time direction $\tau$ with a given period $\beta^{-1}$ to avoid a singularity at the horizon. For the extreme hole we find that we can compactify $\tau$ with {\it any} period; since the horizon is infinitely far away we do not get a singularity from a `wrong' choice of period. 

On the other hand  string theory has considered these extremal holes from a microscopic viewpoint, and used an indirect argument (going to weak coupling) to count the states of the system. The entropy agrees exactly with the area entropy $S={A\over 4}$ \cite{sv}. 

So the question arises: can we find a way to reconcile these two results on the entropy of extremal holes?

\subsection{Entropy and the throat geometry}

We do not know how to construct all the capped geometries for the 3-charge system, but there are several pieces of evidence for the fact that the `cap' structure of 3-charge and 4-charge systems is analogous to that in the 2-charge case. In \cite{gms} capped geometries were constructed for a special family of states, and the depth of the throat upto the cap gave an energy gap that exactly agrees with the gap in the CFT.  We can `squeeze' the 3-charge hole by compactifying a transverse direction; this gives a 3-charge extremal system in 3+1 noncompact dimensions. This system is completely analogous to the 2-charge system in 4+1 dimensions; the throat narrows down as we go to $r=0$. An indirect argument in \cite{fuzzhigh} showed that for this system of 3-charges in 3+1 dimensions the throat depths for the generic state were consistent with the energy gaps in the CFT. Large families of capped geometries have been found in \cite{bena,bena2,gimon,gimon2,gs}, and the maximal throat depth for systems with $U(1)\times U(1)$ symmetry was related to the CFT energy gap in \cite{bena2}.

 The metric  the naive 3-charge geometry  has the form
 \be
 ds^2=(1+{Q\over r^2})^{-1} [-dt^2+dy^2]+(1+{Q\over r^2})[{dr^2}+r^2d\Omega_3^2]+dy^2+dz_adz_a
 \label{naive3}
 \ee
 This naive geometry is pictured in fig.\ref{first}(a), while the fuzzball  picture of `caps' is fig.\ref{first}(c).

We now see a possible reconcilation between the  two results mentioned above for the entropy of extremal holes. The classical calculation which obtains $S=0$ considers a completely uniform infinite throat, and indeed in the fuzzball picture there are no states at the end of the infinite throat. The string calculation, on the other hand,  started with small coupling $g$, where the string bound state was a small quantum object. As we increase $g$, the geometry picks up a deeper and deeper throat, but at no point does the throat become strictly infinite. To catch the states that have the entropy $S_{micro}$ we must stop at a large but finite distance down the throat, where the generic caps appear. For each string state the geometry at the dashed boundary of fig.\ref{first}(c) has some distortion from spherical symmetry. If we insisit on complete spherical symmetry we get the infinite throat, but this does not correspond to any state in the fuzzball picture, in agreement with the classical result.

\section{Quantum corrections and the nature of the fuzzball}\label{nature}
\setcounter{equation}{0}

We have looked at the `size' of the fuzzball for a sub-ensemble of states of the 2-charge system, and noted that its surface area satisfies a Bekenstein type relation with respect to its entropy. At this point we should check that quantum corrections do not make our observation meaningless; i.e., we should look at the sources of corrections to the classical geometries that we have used and check that their effects do not modify the essential properties of our states.

The generic state is of course not well described by a classical solution -- it is a quantum `fuzzball'. There are two sources of quantum effects:

\bigskip

(a) {\it Low occupation numbers $\langle m_k\rangle$:}\quad The classical solutions (\ref{qsix}) were generated by starting with a classical profile $\vec F(v)$ of the string. Fourier transforming the $F_i(v)$ in $v$ we arrive at a discrete set of harmonics, each of which is quantized as a harmonic oscillator.\footnote{This way of understanding states was true in the weak coupling computation reviewed in section \ref{genericsize}, and since the states are BPS it will also give us the count at strong coupling.  For a direct computation of the phase space structure at strong coupling, see \cite{rychkov}.} Consider the oscillator for a fourier mode $k$. If the occupation number $m$ of this mode is $m\gg 1$, then we get the same essential physics whether we look at energy eigenstates $|m\rangle$ or coherent states\footnote{More detailed work with coherent states/energy eigenstates can be found in \cite{skenderis}.}
\be
|\psi>=e^{-{|\alpha|^2\over 2}}e^{\alpha a_k^\dagger}|0\rangle, ~~~~~~|\alpha|^2= m
\ee
But in the generic state the  oscillators have occupation numbers
$\langle m\rangle \sim 1$ (eq.(\ref{unity})). So coherent states are not a good approximation, and all we have is a collection of eigenstates with  low occupaption numbers for a large number of different harmonics $k$. It is not hard to see  that the transverse {\it size} of the state depends on  the mean harmonic number $\bar k$, while the {\it fluctuations} depend on the mean occupaption number per harmonic $\langle m\rangle$. So the generic state has fluctuations of order unity, but we can estimate its size by using a classical state where all the energy is placed in a one harmonic of order $\bar k$.

\bigskip

(b) {\it String 1-loop  corrections:}\quad We can have stringy effects that correct the state. The leading terms of this kind give $R^4$ terms in the low energy effective action. In \cite{fuzzhigh} it was noted that this term arises at one-string-loop order  where an NS1 winding around the $S^1$ propagates around the loop. For the case of ${\cal M}=$ K3 compactifications we get an effective $R^2$ term in the noncompact directions, and this becomes order unity near the boundary of our generic state. It was shown in \cite{fuzzhigh} that this term remains bounded even though the $S^1$ shrinks to zero size; this happens because the $S^1$ is a contractible loop in the full `capped' geometry. In the case ${\cal M}=T^4$, we get just an $R^4$ term in the noncompact directions.  As we will see,  the ratio of this $R^4$ term to the leading $R$ term is small, and we can explore a large range of the parameter $M$ in defining our sub-ensemble without the $R^4$ corrections becoming significant at the boundary.

\subsection{The string 1-loop terms}

As noted in \cite{fuzzhigh}, the $R^4$ terms in the effective action relevant to our system arise from string 1-loop effects, where a NS1 winding around the $y'$ circle propagates around the loop. Let us write this term for IIB string theory, in the string frame, in the 8+1 dimensional space obtained by reduction on the $S^1$ along $y'$. We get (the notation is explained in \cite{fuzzhigh})
 \be
 S_{1-loop}={1\over 8 G_N^{(10)}}g^2 {\pi^2\over 9\cdot 2^6}\int d^9 x \sqrt{-g_s}[\hat R(t_8t_8+{1\over 4}\epsilon_9\epsilon_9)R_s^4
 +{1\over \hat R}(t_8t_8-{1\over 4}\epsilon_9\epsilon_9)R_s^4]
 \ee
 Here $\hat R$ is the radius of the $S^1$ in the direction $y'$.  This circle becomes small at the fuzzball boundary, so the term in ${1\over \hat R}$ is the one of interest to us. The leading order term in the action is
 \be
 S={1\over 16\pi G_N^{(10)}}\int d^9 x \sqrt{-g_s} (2\pi \hat R) e^{-2\phi}R_s
\ee
Assuming $\phi\approx 0$,  the ratio of the relevant part of the 1-loop correction and the leading term is
\be
\nu\sim {g^2\over \hat R^2} R_s^3
\label{pthree}
\ee

Now let us consider the D1-D5 system and compute the value of $\nu$ at the boundary of the generic fuzzball. This boundary is at $r'=b={g'\over \sqrt{V'} R'}$ (eq. (\ref{bvalue})), and we will use the naive metric (\ref{naived}) to compute the factors in $\nu$. The length of the compact $y'$ circle was computed in (\ref{length1}), so we have
\be
\hat R^2\sim R'^2 {b^2\over (Q'_1Q'_5)^{1\over 2}}
\label{pone}
\ee
The geometry (\ref{naived}) is locally $AdS_3\times S^3\times T^4$ in the throat, so the curvature scale can be read off from the radius of the $S^3$. We have 
\be
R_s\sim {1\over (Q'_1Q'_5)^{1\over 2}}
\label{ptwo}
\ee
Substituting (\ref{pone}),(\ref{ptwo}) in (\ref{pthree}), we find
\be
\nu\sim {V'^2\over g'^2} {1\over n'_1n'_5}
\label{pfour}
\ee
To study the black hole properties of the 2-charge system we keep the moduli $V', R', g'$ fixed and take the limit $n'_1, n'_5 \gg 1$. We observe that 
\be
\nu\sim {1\over n'_1n'_5} <<1
\ee
so the 1-loop terms do not give a significant correction at the boundary of the fuzzball region. 

Now let us look at the boundary of the sub-ensemble with mean harmonic order $\bar k_M=M\bar k$ (eq.(\ref{knew})). From (\ref{bvaluenew}) we see that the value of $b$ is now smaller by a factor $M$. Calling $\hat R_M$ the new radius of the $y'$ circle, we have
\be
\hat R_M^2={1\over M^2} \hat R^2
\ee
while $R_s$ remains unchanged (we still have the same geometry $AdS_3\times S^3\times T^4$). Thus we find
\be
\nu\sim M^2{V'^2\over g'^2} {1\over n'_1n'_5}\sim M^2{1\over n'_1n'_5}
\ee
where in the last step we have again ignored the moduli. We therefore see that in computing the area of the boundary of the cap region we can ignore the 1-loop effects as long as we take
\be
1<<M<<(n'_1n'_5)^{1\over 2}
\ee
This allows a large range for $M$, and thus allows us to consider a large set of sub-ensembles.

In principle this is what we needed to check; if the $R^4$ corrections
do not significantly affect the region at the boundary, then the area of this boundary has been correctly estimated. We do not actually need to know the details of the state inside the fuzzball. But since we have set up the tools to estimate length and curvature scales, we try to obtain a rough picture of the fuzzball interior by extending these estimates. 

\subsection{The interior of the fuzzball for ${\cal M}=T^4$}

As noted at the start of this section, there are two kinds of quantum effects that we must consider. The quantum fluctuations of type (a) are always there for the generic state, and gives  quantum fluctuations of order unity, but this is a simple effect following from the basic quantum mechanics of harmonic oscillator wavefunctions. We would thus like to note it and move on to exploring effect (b). To do this we assume that the occupation numbers of harmonics have been chosen to be somewhat larger than unity  so that the quantum fluctuations of type (a) are suppressed. Given this assumption, we would like to get a general picture of the geometry in the interior of the boundary, and then see what the corrections of type (b) do.

Let us work in the D1-D5 frame. For simplicity we will set all moduli to be order unity
\be
R'\sim 1, ~~~V'\sim 1, ~~~g'\sim 1
\label{choice}
\ee
Thus string length, 10-d planck length and 5-d planck length are all of the same order, and this length has been set to unity
\be
l_p\sim l_s\sim 1
\ee
 We let the charges be equal
\be
n'_1=n'_5=n\gg 1
\ee
We have
\be
Q'_1\sim Q'_5\sim n
\ee
Note that the coordinate size of the generic state is given by 
\be
|\Delta \vec x'|=b\sim 1
\ee
Using the metric (\ref{qsix}), we observe the following

\bigskip

(i) The length of the circle $S^1$ at $r\sim b$ is
\be
L_y\sim ({b^4\over \sqrt{Q'_1Q'_5}})^{1\over 4}\sim n^{-{1\over 2}}
\label{ll}
\ee
Note that this is small; its smallness is reposnible for the fact that $R^4$ terms can become significant for K3 compactifications \cite{fuzzhigh}.

\bigskip

(ii) The proper length of the diameter of the region in $\vec x'$ space enclosed by the boundary is
 \be
D\sim ({Q'_1Q'_5\over b^4})^{1\over 4} b\sim n^{1\over 2}
\label{dvalue}
\ee
Note that the size of our `horizon' is therefore much larger than string or planck length, and thus the 2-charge system should be thought of as a good model for black holes.

\bigskip

(iii) Now consider the picture of the `ball' drawn in fig.\ref{intro}(a). The path of the NS1 traces out a convoluted curve inside this ball, and we wish to understand some properties of this curve. For example, we would like to know the typical distance between one segment of this curve and the closest neighboring segment. We do this as follows.

\begin{figure}[htbp] 
  \hskip 0.7 truein \includegraphics[width=3.5in]{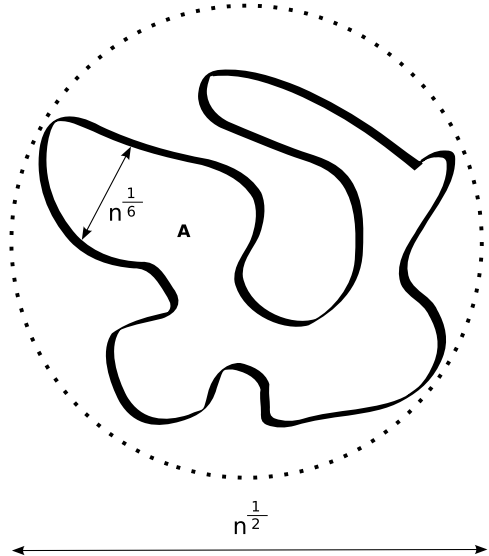} 
   \caption{Length scales in the generic fuzzball. The curve has unit thickness, and the other length scales are larger by powers of $n$. In the generic state the location of the thick curve has quantum fluctuations that are large but these have been suppressed (by choosing high occupation numbers of modes) to enable an estimate of the string 1-loop effects. }
   \label{ball}
\end{figure}

Look at the top sketch in fig.\ref{intro}(a). We know that the generic state has wavenumber $\bar k\sim\sqrt{n_	1n_p}\sim n$. Thus there are $\sim n$ oscillations of this curve on the string. Each oscillation, roughly speaking, makes the string traverse once across the ball.  We draw the 4-dimensional space $\vec x'$   in fig.\ref{ball}. Let us make a 3-dimensional section through the center of this 4-d region. There will be $\sim n$ intersections of the curve with this plane. Since the proper diameter of this ball  is $\sim n^{1\over 2}$ (eq. (\ref{dvalue})), these intersections occur in a 3-dimensional area 
\be
A_3\sim  n^{3\over 2}
\ee
Thus the density of intersection points in this plane is
\be
\sigma\sim {n\over A_3}\sim n^{-{1\over 2}}
\ee
and the typical proper distance between points on this plane will be
\be
d\sim {1\over \sigma^{1\over 3}}\sim n^{1\over 6}
\label{dd}
\ee
Thus we see that the strands of the curve do not come within planck or string distance of each other, but stay quite far from each other in generic configurations.

\b

(iv) Let us note the `thickness' of the curve traced out by the string. We are working in the D1-D5 frame. A shown in \cite{lmm}, each  point on the curve is not
a singularity but the center of a KK monopole. The curvature length scale of this monopole is set by the length of the fibred circle, which is the $y'$ circle $S^1$, and this has length $\sim n^{-{1\over 2}}$ (eq. (\ref{ll})). This is very small (much less than planck length). This situation is not a surprise, since it is exactly what happens with any brane in string theory: there can be a high curvature in a planck region around the brane, and this is just included in the description of the brane. The KK monopole tube is dualizable (locally) to any other brane in string theory, and so we will leave a planck sized tube around our curve as a region where we will not trust the classical metric.

\bigskip

(v) Now let us look at the harmonic functions $H_1, H_5, A_i$ that give the metric (\ref{qsix}). All these functions have a similar structure, so let us look at $H_1$. First let us go close to  a point on the curve $\vec F(v)$, say at a point $v=v_0$. Near this point, the curve looks like a straight line; let $\rho$ be the coordinate distance from this line in the coordinates $\vec x'$ that we have used on the noncompact space for the D1-D5 system. If we are sufficiently close to the curve $\vec F(v)$ then we can replace the curve by the line, and we find\footnote{In \cite{fuzzball} it is shown that $H^{-1}\approx {Q'_5\pi\over \mu^2 L_T |\dot{\vec F}| } {1\over \rho}$. But with our choice of moduli (\ref{choice}), we have $\mu\sim 1$, $Q'_5\sim n$, $L_T\sim n$, and from $Q'_1\sim Q'_5$ we know that $\dot{\vec F}\sim 1$. Thus we get $H^{-1}\sim {1\over \rho}$.}
\be
H_1\sim {1\over \rho}
\label{rho}
\ee
Now consider a point, like point A in fig.\ref{ball}(a),  which is {\it not} close to any point on the curve $\vec F(v)$. There is a contribution to $H_1$ from segments of  $\vec F(v)$ that are close to the point A, and a contribution from segments further away. The closer segments are fewer, but for them ${1\over |\vec x'-\vec F(v)|}$ is larger; the further segments are more numerous, but ${1\over |\vec x'-\vec F(v)|}$
is smaller. Which contribution is more important?

A simple estimate tells us that the `far away' contribution is larger, but note that it also changes smoothly with position, since it comes from strands that are far away from the point A. Let us call it $H_1^{smooth}$. The `nearby' contribution is smaller, but it fluctuates on the scale of the distance between neightbouring strands, which we have seen is $\sim n^{1\over 6}$. We call it $H_1^{fluc}$. We find
\bea
H_1&=&H_1^{smooth}+H_1^{fluc}\nn
H_1^{smooth}&\sim& n\nn
H_1^{fluc}&\sim &\sim n^{1\over 3}
\eea
(To estimate $H_1^{smooth}$ we note that $Q\sim n$ and $|\vec x'-\mu \vec F|\sim 1$ in (\ref{functionsqq}). To estimate $H_1^{fluc}$ we have noted that the {\it coordinate} distance $\rho$  from point A to the nearest strand is the proper distance (\ref{dd}) divided by $n^{1\over 2}$; here the factor $n^{1\over 2}$   is the ratio between coordinate and proper distances in $\vec x'$ space. We then substituted this $\rho$ in (\ref{rho}).)

The function $H_5$ has the same behavior, and in $A_i$ we find that the smooth part cancels to zero (since the orientation of the vector $\vec A$ changes randomly along the curve $\vec F(v)$) while the fluctuating part has the same order as $H_1^{fluc}$. Since the functions $H_1, H_5, A_i$ appear as coefficients in the metric (\ref{qsix}), we find that the metric in the space $\vec x'$ (schematically called $g$) itself has a smooth part and a fluctuating part
\be
g=g^{smooth}+g^{fluc}
\ee
with
\be
h\equiv {g^{fluc}\over g}\sim n^{-{2\over 3}}
\label{hhh}
\ee

\b

(vi) Finally, we use the length scales found above to examine the $R^4$ effects. The smooth part of the metric $g^{smooth}$ varies over a distance of the order of the radius $n^{1\over 2}$ of the ball. Thus it gives a curvature $R_s\sim n^{-1}$. Thus for $\nu$ we have
\be
\nu\sim {n^{-3}\over n^{-1}}\sim {1\over n^2} <<1
\ee
where we have used (\ref{ll}) for $\hat R^2$. The fluctuating part varies on a smaller distance scale $d$, but $g^{fluc}$ is also small. Thus we get a curvature
\be
R^{fluc}_s\sim {h\over d^2}\sim n^{-1}
\ee
which again gives $\nu\sim {1\over n^2}<<1$.

\b

To summarize, we get the picture sketched in fig.\ref{ball}. All lengths are measured in units of planck length. We have a ball of diameter $\sim n^{1\over 2}$. In this balls a curve winds around; the quantum  `thickness' of this curve is unity. The strands of this curve do not come too close to each other; the typical distance to the nearest neighbor is $\sim n^{1\over 6}$. The $R^4$ corrections are small. The ordinary quantum fluctuations of type (a) are always present, and these will cause the curve to fluctuate, and be in a general quantum wavefunction.

\b\b

\section{Discussion}\label{discussion}
\setcounter{equation}{0}

We have looked at some subsets of states of the 2-charge extremal system, and found that in each case the boundary of the `fuzzball' satisfies a Bekenstein type relation with the entropy inside the ball. This suggests that the system requires a minimum sized region to contain its phase space, and black holes `pack states' into these minimal regions. Let us note some ways in which phase space considerations can impact the puzzles arising with black holes.

\subsection{The quantum nature of the `cap'}

In the classical treatment of a black hole we do not see the states of the hole (`black holes have no hair'). Thus we miss an essential feature of the hole, since the entropy is actually very large. Extremal holes may be the easiest to understand. We can start with the brane bound state at weak coupling $g$; this state has a large degeneracy. Since the states are BPS, we can follow them to strong coupling. The geometry develops a throat, and the throat gets deeper with increasing $g$, but by continuity in $g$ we do not expect  at any given $g$ to have an infinite throat with a horizon at the end. The `caps' at the end of the throat have a structure that reflects the particular microstate. In the 2-charge D1-D5 system the `true charges' D1,D5 give rise to a `dipole ring' which takes the form of a KK monopole times a $S^1$. Different shapes of this ring give different classical geometries. Quantum fluctuations (which we reviewed in this paper)  make the generic state a `fuzzball' with the same size as observed in the classical geometries.

For three and four charge systems the dipole charges give more complex structures to the cap, and large families of classical solutions have been found starting from these dipole charges \cite{bena,bena2,gimon,gimon2,gs}.  In the 4-charge solutions the dipoles are points in the noncompact space. As we go to more and more generic states, the number of these dipole charges can be expected to grow, and we will have a vast collection of `branes at angles' in the `cap' region \cite{gimon2}.\footnote{An elegant and detailed analysis of  bound states of branes and black holes can be found in \cite{denef}.} We can understand all D-branes as arising from tachyon condensation in other sets of branes and anti-branes; this is the `K-theory' picture of D-branes.\footnote{An excellent review can be found in \cite{evslin}.} Thus we can visualize the generic `cap' state as a wavefunction in this very complex potential made from an arbitrary number of branes. If the tachyon condenses to its minimum in a given charge sector then we see the BPS dipole branes observed in the classical solutions \cite{gimon2}, but more generally the system will just have a complicated wavefunction in this multi-dimensional tachyon potential. The entropy of non-BPS holes can be understood in terms of brane-antibrane pairs \cite{hms}, so for these holes we expect a similar potential with a large number of local extrema; the slow decay by Hawking radiation would then correspond to the system drifting to its true ground state. If this picture of black hole states is correct, then we must understand the entropy of the hole in terms of the number of metastable states possible in the  tachyon potential arising from an arbitrary number of brane pairs.\footnote{ For non-BPS system we can also imagine making the analogues of classical states where all quanta are placed in just a few modes; such states were constructed in \cite{ross}.} 

Put another way, if we are given an energy $E$, then the maximal entropy is not obtained by converting this energy into massless quanta, but rather by using it to create brane and anti-branes, which is a system with a large phase space, and gives a high degeneracy of states.

\subsection{Phase space and the classical geometry}

The traditional puzzle with black holes has been the following. We have a given classical geometry. There seem to be no large corrections at the horizon. Why should string theory change anything?

Let us address this question in the context of extremal holes. The classical geometry has an infinite throat. Nothing seems to be wrong with this geometry by itself. But the recent work on fuzzballs shows us that there are a large number of {\it other} states with the same mass and charge, where the throat behaves differently after a certain depth. Why are these states relevant? 

\begin{figure}[htbp] 
   \centering
   \includegraphics[width=4in]{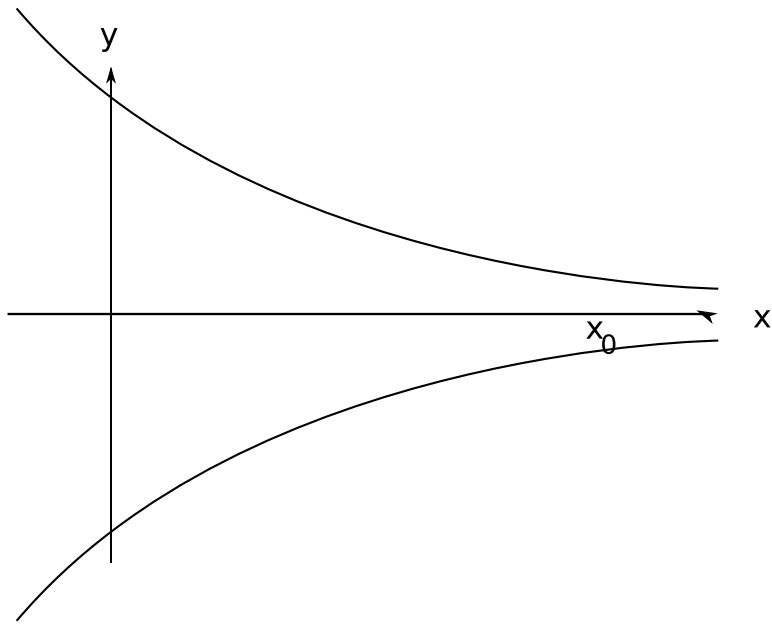} 
   \caption{The $x$ axis denotes one direction of a phase space; this would appear to have an infinite number of unit cells beyond the point $x_0$. The $y$ axis denotes another direction of phase space; since this narrows to the right there are only a finite number of states to the right of $x_0$ in the full phase space.}
   \label{phase}
\end{figure}

If we quantize the system, we have one state per unit cell in phase space. But to find this unit cell, we first have to understand all the directions of this phase space. To illustrate this point schematically, consider a 1-dimensional space (drawn as the horizontal line in fig.\ref{phase}). Since this line is infinite, it would appear that there will be an infinite number of unit cells along its length. But suppose there was another direction $y$ to this phase space, and that the volume of phase space (available at a given energy) narrows down to the right as shown in the figure. Now we see that beyond some point $x_0$ there is very little phase space left, and this available phase space might be able to hold only a finite number of states.

It may be possible to see such a situation in the 2-charge extremal system. First let us split our D1-D5 system with charges $n'_1, n'_5$ into two smaller systems,  with charges  $n^{(1)}_1, n^{(1)}_5$ and $n^{(2)}_1,n^{(2)}_5$ respectively. The relative motion of these two objects is described by a moduli space metric, which we can find from \cite{jeremy} by restricting their 3-charge answer to the case where only two charges are present. We find (for small separations $r$)
\be
dS^2={(n^{(1)}_1n^{(2)}_5+n^{(1)}_5n^{(2)}_1)\over r^2}(dr^2+r^2d\Omega_3^2)=
(n^{(1)}_1n^{(2)}_5+n^{(1)}_5n^{(2)}_1)[{dr^2\over r^2}+d\Omega_3^2]
\label{je}
\ee
We see that the angular part of this moduli space has a constant volume, while the radial part is an infinite line. (We set $z=\log r$ and find the moduli space metric $\sim dz^2$ for $-\infty<z<\infty$.) This suggests an infinite number of quantum states.\footnote{Note that this is what we call an `unbound system' \cite{fuzzball}. But when $r\r 0$ we have a single throat that branches into two throats after a great depth, so we are quantizing some degrees of freedom  of the geometry of the 2-charge throat.} 

But we know that there are no throats longer than a maximal depth in the correct quantization of the 2-charge system, and there are a finite number of states overall. In the CFT dual, the longest throat corresponds to a single `component string' of length  $2\pi R'n'_1n'_5$ \cite{lm4}. In the NS1-P duality frame, the longest throat corresponds to having a single vibration in the harmonic $n_1n_p$ carrying all the momentum charge. We cannot go to higher harmonics of vibration, since then we will need a fractional number of quanta to carry the given momentum charge. This fractional number of quanta can be thought of as an excitation that was allotted a fractional unit of phase space to exist; such an excitation is not possible.

The above considerations were arrived at using the CFT dual, but the phase space of the D1-D5 system can also be quantized directly in the gravity description, and the same measure on phase space is obtained \cite{rychkov}. Thus we find that when we quantize the full phase space of the system, rather than just one direction like in (\ref{je}), then all throats longer than a given depth fall into a single wavefunction, and it does not make sense to explore the throat to arbitrary depths in the manner suggested by the classical metric.

We hope to return to these issues in the future.

\section*{Acknowledgements}

I am grateful to Stefano Giusto for help with many aspects of this paper. I would  like to thank Borundev Chowdhury and Jeremy Michelson for helpful comments. I would also like to thank Iosif Bena, J. de Boer, Eric Gimon, Tom Levi and Nick Warner for sharing their thoughts on related projects.  This work was supported in part by DOE grant DE-FG02-91ER-40690.

\end{document}